\documentclass[usenatbib,useAMS,preprint]{mn2e}
\usepackage{graphicx}

\def\gapprox{\lower.4ex\hbox{$\;\buildrel >\over{\scriptstyle\sim}\;$}}
\def\lapprox{\lower.4ex\hbox{$\;\buildrel <\over{\scriptstyle\sim}\;$}}

\def\bk{\mbox{\boldmath $k$}}

\def\be{\mbox{\boldmath $e$}}

\def\bp{\mbox{\boldmath $p$}}

\def\bB{\mbox{\boldmath $B$}}
\def\bE{\mbox{\boldmath $E$}}
\def\bV{\mbox{\boldmath $V$}}

\def\ICS{{_{\rm ICS}}}
\def\NS{{_{\rm NS}}}

\title[Pulsar transient radio emission]{Pulsar radiation belts and transient radio emission}
 
\author[Luo, et al]
      {Qinghuan Luo and Don Melrose\\
        School of Physics, The University of Sydney, NSW 2006, Australia\\
}
                                                                                                            
\date{
          --- Received
         in original form February, 2007
        }
                                                                                                            
\pubyear{2006}
                                                                                                            
\begin{document}
                                                                                                            
\maketitle
                                                                                                            
\begin{abstract}
It is proposed that radiation belts similar to the ones in the planetary 
magnetosphere can exist for a pulsar with a relatively long period and a 
strong magnetic field. In the belts located in the closed
field line region near the light cylinder relativistic pairs
are trapped and maintained at a density substantially higher than the 
local Goldreich-Julian corotation density. The trapped plasma can be supplied
and replenished by either direct injection of relativistic pairs from acceleration
of externally-supplied particles in a dormant outer gap or {\it in situ} ionization 
of the accreted neutral material in the trapping region. The radiation belts can be disrupted by 
waves that are excited in the region as the result of plasma instabilities or emitted from the surface 
due to starquakes or stellar oscillations. The disruption can cause an intermittent 
particle precipitation toward the star producing radio bursts. It is suggested that 
such bursts may be seen as rotating radio transients (RRATs).
\end{abstract}
                                                                                                            

\section{Introduction}           
\label{sect:intro}

The recent discovery of a new class of
radio transient sources, known as rotating radio transients (RRATS)
\citep{metal06}, suggests that radio transient phenomena may be 
quite common for typical pulsars. About 11 such sources have been detected so far;
they are characterized by short outbursts of a typical from 2 to 30 millisecond duration
with an average interval between two consective bursts from a few minutes
to a few hours. Ten RRATs have their periodicities determined and three of them 
have their period derivatives identified. Observations appear to suggest that these objects 
are part of the normal pulsar population, with periods of a few seconds, within the period
range of typical radio pulsars, and magnetic fields up to $10^{10}\,\rm T$, 
close to a lower limit of the magnetic field of the known magnetars.
No transient X-rays or optical counter part has been observed, though 
thermal X-ray emission was detected from one RRAT, which appears to 
be similar to X-ray dim isolated neutron stars 
(XDINS) \citep{retal06,petal06}. 

Three RRATs with both periods and period derivatives
identified overlap with normal radio pulsars in the $P$-$\dot{P}$ (pulsar
period vs period derivative) distribution,
which strongly suggests that RRATs may have similar properties to radio pulsars and should
have normal radio emission as well.  However, regular radio pulses have not been detected
from any of these sources, suggesting that either (1) these radio bursts are distinct
from the normal radio emission in aspects of emission geometry or processes or (2)
normal radio pulses are too weak to be detectable \citep{wetal06a}.
For case (1), one could postulate, for example, that radio emission 
jumps between two alternative beaming direction, one of which intersects
the Earth. For case (2), RRATs may be considered as extreme pulses and
the `missing' normal radio pulses can be detected with a sufficiently
long observing time \citep{wetal06a}. \citet{wetal06a,wetal06b} specifically
considered PSR B0656+14, which is a nearby pulsar that has 
pulsed X-rays and intense bursty radio pulses similar to RRATs.
They suggested that if this pulsar were at a distance
similar to that RRATs, one would see the strong radio bursts as RRATs and its
normal radio emission becomes nondetectable. 

If RRATs are indeed pulsars, one has a major difficulty in interpretation of 
the observed intense radio bursts on the basis of conventional polar cap models because
most of RRATs have long periods and pair production is not effective.
\citet{cs06} proposed that the bursty emission by RRATs is due to 
circumpulsar asteroids randomly straying into the magnetosphere; the
neutral material is evaporated and ionized, leading to a sudden downward flow of 
charged particles in the polar region, which in turn ignites a transient pair cascade toward
the PC. The low-mass disk hypothesis has some observational support from the recent
discovery of the fall-back disk of a magnetar, cf. \citet{wck06}. One main ingredient of 
the Cordes \& Shannon's model is that a cascade is assumed to be due to acceleration in the outer
gap. However, RRATs are in the regime in the $P$-$\dot{P}$ distribution
where the outer gap is inactive for pair production (through photon-photon
collision) except for the case of a nearly orthogonal inclination angle where the
outer gap may form close to the PC and pair production through a single photon decay
in the magnetic field is important. In this latter case,
the cascade has to be close to the PC and backward emission from the cascade
may not be able to propagate freely through the intervening plasma near
the star in the closed field line region (CFLR)
where induced three-wave interactions are expected to be strong \citep{lm06}.

In this paper, we propose that typical pulsars may have radiation belts 
in  the CFLR where relativistic plasmas are trapped.
The trapping regions bear many similarities to the Earth's radiation belts. 
For example, they are similarly subject to various low-frequency disturbances that disrupt 
the trapped plasma causing intense precipitation. While in the case of the Earth's van Allen belts,
which are subject to disruption by geomagnetic storms due to the solar wind, 
the proposed main disturbances to the
pulsar's radiation belts are low-frequency Alfv\'en waves generated from the star's surface as a
result of stellar oscillations \citep{metal88} or shear waves in the neutron star's crust
due to starquakes \citep{betal89}. It is suggested
here that transient radio emission similar to RRATs can be generated as a result of
catastrophic disruption of the trapping region and that the coherent emission is produced
by particles precipitating toward the star. 

The density of the trapped plasma can be maintained at a much higher value 
than the local GJ density. In the trapping region the magnetic 
field is relatively weak and the synchrotron decay time is long so that the 
plasma can be replenished, leading to a build-up in the density.
The major sources of the trapped plasma can be particle acceleration in a dormant
outer gap or ablation and ionization of the accreted neutral matter.
We assume that low-level accretion of neutral grains from a dust disk or from
the ISM occurs and that the accreted matter is ionized inside the LC
feeding charged particles directly to both the open field line region
(OFLR) \citep{c85,rc88,cs06} and CFLR \citep{cs06}. Note that there were 
also previous discussions on accretion of neutral material from the ISM and 
possible effects on radio emission by ionization of the neutral matter in 
the pulsar magnetosphere \citep{t77,w79}. Charged particles created from
destruction and ionization of neutral material in the CFLR can accumulate and
be trapped in the region. Charged particles created in the OFLR can be
accelerated in the outer gap and inject relativistic pairs into
the trapping region. High-energy gamma rays emitted by
the accelerated particles produce pairs in the trapping region
on the thermal radiation emitted from the surface. The pair production rate
is rather slow, but provided that it exceeds the loss rate, 
it can accumulate plasma near the LC in the CFLR.
Note that reignition of a latent outer gap was discussed by \citet{rc88}
in the context of gamma-ray bursts. However, they considered nearly aligned rotators
with a period much shorter than that considered here. In their model,
injection of externally-supplied particles in the outer gap
is assumed to lead to efficient pair creation, a case that does not apply
to RRATs or magnetars.

In Sec. 2, it is argued that plasma trapping regions
similar to the Earth's radiation belts can exist in the CFLR of typical pulsars
with a long period. Injection of charged particles to the trapping region
due to latent particle acceleration in an outer gap or direct ionization of accreted 
neutral dust grains is discussed in Sec. 3 and 4.
Sec. 5 discusses the stability of the pulsar radiation belts and possible
mechanisms for particle precipitation. Application to RRATs is discussed
in Sec. 6.

\section{Radiation belts}

It is argued here that radiation belts similar to that in planetary
magnetospheres may exist in pulsar magnetospheres with relatively weak magnetic
fields at the LC. 

\subsection{Magnetic mirror}

Particles injected with nonzero perpendicular
momenta can be trapped in the CFLR due to the magnetic mirror
effect. For a particle with a perpendicular momentum $p_\perp$, the first 
adiabatic invariant can be expressed as $p^2_\perp/B={\rm const}$. In this section,
we ignore the loss of perpendicular energy due to cyclotron decay, which can be important
for pulsars and is considered in Sec. 2.2. From a quantum mechanical view point, this corresponds to
a particle remaining in the same Landau level. The magnetic field in the CFLR
forms a `magnetic bottle' and particles injected at a radial height $\eta_{i}<1$
($\eta=r/R_{LC}$ is the radius in unit of $R_{LC}=cP/2\pi$)
can be trapped at $\eta<1$ provided that their perpendicular momenta satisfy the condition:
\begin{equation}
{p_\perp\over p}>\left({\eta\over\eta_{i}}\right)^{3/2},
\label{eq:mirror}
\end{equation}
where $p=(p^2_\perp+p^2_\parallel)^{1/2}$ with $p_\parallel$ the parallel momentum.
For example, particles injected with $p_\perp/p_\parallel>0.35$ at the LC
($\eta_{i}=1$) can be trapped; they bounce back and forth between two opposite hemispheres
in the CLFR above $\eta=0.5$. The right-hand side can be written into the
form $\sin\alpha_c=(\eta/\eta_{i})^{3/2}$, where the angle $\alpha_c$ defines a conal surface
in the momentum space,  referred to as the loss cone at $\eta$. Particles
with a pitch angle smaller than $\alpha_c$ (inside the cone) pass through the point $\eta$
and those with a pitch angle larger than $\alpha_c$ (outside the cone) are reflected above $\eta$.
For particles moving along the last closed field lines, the typical bounce
time is (cf. Appendix)
\begin{equation}
\tau_b\approx {2\over\pi}P,
\end{equation}
which is not sensitive to the particle's initial pitch angle $\alpha_i$
at the injection but required to satisfy the condition
\begin{equation}
\alpha_i\geq \left({R_0\over R_{LC}}\right)^3\approx
10^{-12}\left({P\over2\,{\rm s}}\right)^{-3},
\label{eq:ai}
\end{equation}
i.e., the reflection radius must be above the stellar radius.
Eq (\ref{eq:ai}) is much smaller than the minimum loss-cone angle
of the van Allen belts in the Earth's magnetosphere, which is a few degrees.
However, it is shown in Sec 2.2 that the minimum loss-cone angle determmined
by cylcotron decay can be much larger than (\ref{eq:ai}).

Since the dipole magnetic field is inhomogeneous, the guiding centers of trapped particles
undergo both a gradient drift and curvature drift across the field lines \citep{nt60}.
These drifts cause particles to circulate around the star forming a ring current,
similar to the ring current in the Earth's radiation belts. The current gives
rise to perturbations to the dipole field at $\eta=1$:
\begin{equation}
{\delta B\over B_L}\sim {N_Lm_ec^2\gamma\over 2U_{BL}}
\sim 10^{-6},
\label{eq:dB}
\end{equation}
where $U_{BL}=B^2_L/\mu_0$ is the magnetic energy density at the LC, $N_L=N_{GJ}\theta^6_d$,
$\theta_d=(R_0/R_{LC})^{1/2}\approx0.012(P/1.5\,{\rm s})^{-1/2}$ is the half-opening angle of the PC,
and $N_{GJ}\approx 2.3\times10^{17}\,(B_s/10^9\,{\rm T})(P/1.5\,{\rm s})^{-1}\,{\rm m}^{-3}$
is the GJ density at the PC where the magnetic field is assumed to be $B_s$.
In practical situations where the plasma density can be much higher than the GJ density (Sec 3), 
the perturbations can exceed the estimate given by (\ref{eq:dB}).
Since the ring current can vary temporally as a result of instabilities in the trapped plasma
(cf. Sec 5), $\delta B$ should be time dependent. The effect of such perturbations on
pulsar electrodynamics and their possible contribution to
pulsar timing noise will be discussed elsewhere.

\subsection{The loss-cone angle}

\begin{figure}
\includegraphics[width=7.8cm]{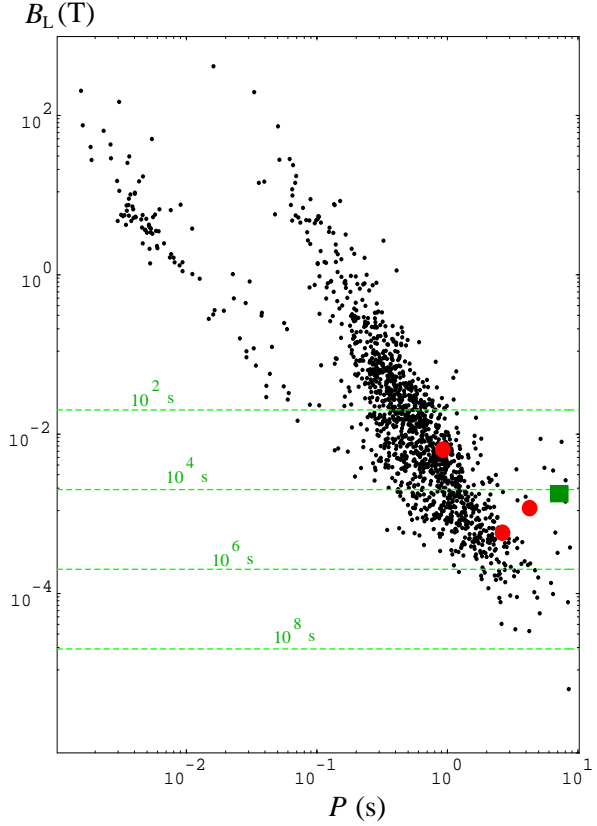}
\caption{Pulsar distribution in $P$-$B_L$ with the decay time $\tau_0$ shown.
The magnetic field at the LC is defined as $B_L=B_s\theta^6_d$. The pulsar sample is
from ATNF pulsar catalogue (http://www.atnf.csiro.au/research/pulsar/psrcat).
The three large shaded dots represent RRATs. The lighter square represents
a typical magnetar.
}
\label{fig:pBL}
\end{figure}

The lifetime of trapped particles is limited by cyclotron decay and
pitch-angle scattering, both of which violate the first invariant. The electron's
synchrotron cooling time, i.e., the time that it loses half of its energy
($m_ec^2\gamma$), is $\tau_s=\tau_0/\gamma$ with
\begin{equation}
\tau_0={3c\over4r_e\Omega^2_e},
\label{eq:tau}
\end{equation}
where $r_e\approx 2.8\times10^{-15}\,\rm m$ is the classical radius of the electron
and $\Omega_e$ is the electron cyclotron frequency. Figure \ref{fig:pBL} shows
a pulsar distribution on $P$-$\dot{P}$ with the time (\ref{eq:tau}) indicated.
Six examples where $\tau_0$ is calculated are listed in Table 1. For fast rotating young pulsars
or millisecond pulsars, (\ref{eq:tau}) is very short $\sim 10^{-6}\,\rm s$,
and for typical pulsars it can be quite long $\sim 10^4-10^5\,\rm s$.
Since in the synchrotron regime, the pitch angle remains constant,
the relevant time that constrains particle trapping is
the cooling time in the limit $p_\perp\to 1$,
$\gamma\to\gamma_c\equiv \sqrt{2}[1+p^2_{\parallel0}/(1+p^2_{\perp0})]^{1/2}$ \citep{letal00},
where $p_{\perp0}$ and $p_{\parallel0}$ are the perpendicular and
parallel components of the particle's dimensionless initial momentum. This gives
$\tau_s\approx \tau_0/\gamma_c$, which has the following numerical form
\begin{eqnarray}
\tau_s\approx 1.4\times10^4 \gamma^{-1}_c
\left({B_s\over10^9\,{\rm T}}\right)^{-2}
\left({P\over 2\,{\rm s}}\right)^{6}\left({\eta\over0.5}\right)^{6}\,{\rm s}.
\end{eqnarray}
One has $\tau_s\sim\tau_0$ if a particle is injected into the trapping
region with a large initial pitch angle $\tan\alpha_0=p_{\perp0}/p_{\parallel0}\approx1$.
However, one has a much shorter cooling time if the initial pitch angle is small $\alpha_0\ll1$.
In the cyclotron regime, the particle radiates at a much slower
rate, relaxing to the ground Landau state on a time
$\tau_s\gamma^2_c \sim \tau_0\gamma_c$.

Equating the cyclotron decay time to the bounce time, one finds the
minimum pitch angle for a particle to be reflected at the mirror point.
This angle is estimated to be
\begin{equation}
\alpha_i\geq 0.04\left({B_s\over10^9\,{\rm T}}\right)^{1/2}
\left({P\over2\,{\rm s}}\right)^{-1/2}.
\label{eq:ai2}
\end{equation}
For particles to remain trapped for time much longer than $\tau_b$, the pitch
angle at injection needs to be much larger than (\ref{eq:ai2}).

Pitch angle scattering by waves can lead to particle diffusion to small pitch angles
and limit the life time of the trapped particles.
Since the decay time decreases rapidly with decreasing altitude according to
$\tau_s\propto \eta^6$, particles scattered to small pitch angles can reach low altitudes
rapidly radiating away their perpendicular energy and eventually reach the star.
For example, particles with $\gamma_c\sim 1$, which are scattered into the loss cone of 
$\alpha_c=0.01$, can travel down to an altitude $\eta=0.05\eta_{i}$ where the decay time is only
$\tau_s\sim (100\, {\rm ms})\eta_{i}$. The sources of the waves can be internal,
i.e., they are generated {\it in situ} through plasma instabilities,
or external, i.e., they are generated outside the region.
Since particles with small pitch angles escape through both
ends of the `magnetic bottle', there is an excess of particles with a
large pitch angle; this gives rise to an ideal condition for instabilities to develop (driven by
an inversion in the particle pitch angle distribution).
Waves generated from instabilities can cause pitch angle diffusion,
further enhancing particle precipitation. In principle, this process can
impose an upper limit to the density of the trapped plasma \citep{kp66}.
However, since the efficiency of pitch-angle scattering depends explicitly on the
wave intensity that in this case is limited to much smaller than the density of
the particle kinetic energy, one needs a relatively high plasma density
for such upper limit to be significant (cf. Sec 5). Alternatively,
pitch-angle scattering can be significant if waves that propagate into
the region can interact with the trapped particles in resonance. This possibility is
discussed in Sec. 5.

\subsection{Sources for the trapped plasma}

Particle loading into the trapping region can be achieved through a process that
is not constrained by the first adiabatic invariant.
We consider two possible sources for the trapped
plasma, both of which involve extrinsic charged particles: (1) pair production
in the trapping region by high-energy gamma rays
emitted by particles accelerated in an outer gap and
(2) direct ionization of neutral material in the thermal radiation field in the CFLR.
Both cases may involve ionization of neutral matter migrating into the magnetosphere
due to accretion of a disk \citep{cs06,l06} or neutral dust grains in the ISM \citep{c85}.
For (1), the outer gap, which is usually dormant for
typical pulsars, is reignited with the supply of charged particles.
Although the gap is rather inefficient for
pair production, it can lead to a build-up of  particles in
the CFLR over a time much longer than the pulsar period (cf. Sec. 4).
For (2), charged particles from the ionization
can be accelerated through either inward (toward the star) cross-field  diffusion
or resonant wave-particle interactions, which is similar to
the particle acceleration in the van Allen belts in the Earth's
magnetosphere \citep{hetal05}.

In principle, high-energy cosmic rays can also inject pairs
into the trapping region. (Note that cosmic rays contribute to trapped
plasmas in the van Allen belts through decay of upward deflected neutrons
produced by high-energy cosmic rays.)  For example,
cosmic protons with a gyroradius larger than $R_{LC}$ can drift into the pulsar magnetosphere.
For a typical pulsar, the proton needs to have a Lorentz factor $\gamma>\Omega_p/\Omega\sim 5\times10^4$,
where $\Omega_p$ is the proton cyclotron frequency at the LC and $\Omega=2\pi/P$.
Ultrarelativistic protons produce pairs on the thermal radiation through
$p+\gamma\to p+e^\pm$. However, it can be shown that such mechanism is not
effective. Assuming a typical proton flux density
similar to that received on the Earth, $F_p\sim 5\times10^{-7}\,{\rm m}^{-2}\,{\rm s}^{-1}$ at
$10^5\,{\rm GeV}$, one finds a pair injection rate
about $10^6\,{\rm s}^{-1}$. It would take
much longer than the Hubble time to fill the magnetosphere
to the GJ density.

\begin{figure}
\includegraphics[width=8cm]{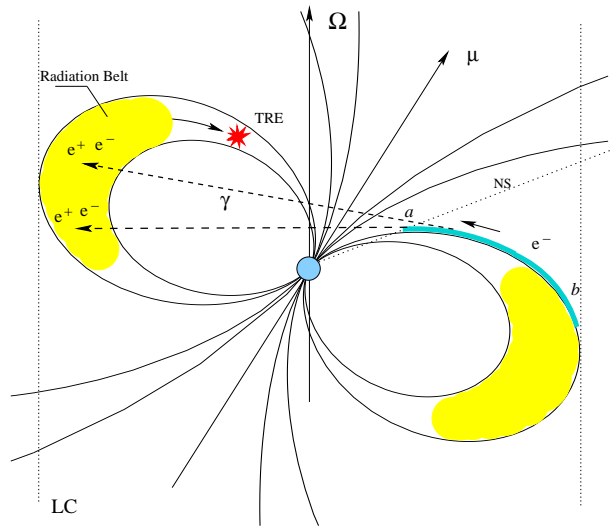}
\caption{Radiation belts (shaded area) and transient radio emission (TRE).
Radio bursts are produced due to particle precipitation toward the star in the 
closed field line zone. The thick line $ab$ along the last open field line is the outer 
gap where electrons are accelerated toward the star. The curvature photon ($\gamma$) 
emitted by these electrons can be converted to
a pair by colliding with thermal photons emitted from the star's surface. Pairs
are then trapped in the shaded region due to the magnetic mirror
effect. Disturbances to the region, such as Alfv\'en waves launched from the star,
can lead to an intense plasma precipitation. As particles move toward the star, coherent 
radio waves can be generated, e.g., through synchrotron/cyclotron maser or two-streaming instability.}
\label{fig:rrat}
\end{figure}

\section{Dormant outer gap}

We show that a dormant outer gap can supply pairs to the trapping regions. 
It is generally thought an outer gap exists between the null surface 
and the boundary surface separating the OFLR and the CFLR
\citep{cetal86,r96,hs02,zetal04,tetal06}. For young, fast rotating pulsars, particle acceleration in
the outer gap can self-sustain a pair cascade, which can transfer a significant fraction of the
gap potential into the secondary particles kinetic energy, which is then radiated at high energies.
As the pulsar period increases, the gap become less efficient in pair production
and thus, for slow pulsars, pair production in the outer gap becomes negligible.
However, provided that there is an external source of particles,
the potential energy of such dormant gap can be tapped and transferred
to plasma kinetic energy trapped in the CFLR.

\subsection{Acceleration in the outer gap}

The accelerating electric field along the field lines in an outer gap can be written as
a fraction of the maximum potential drop $\Phi_m$ across the PC:
\begin{equation}
E_\parallel={\Phi_mw_g\over R_{LC}},
\label{eq:Ep}
\end{equation}
where $w_g\leq 1$ is a parameter characterizing the geometry of the gap and
$\Phi_m=0.5\theta^4_dB_sR_0\approx1.7\times10^{13}\,{\rm V}\,(B_s/10^9\,{\rm T})
(P/2\,{\rm s})^{-2}$. Since pair production can change the gap geometry,
the parameter $w_g$ is usually determined self-consistently by taking into
account the effect on the accelerating potential by pair creation
\citep{tetal06}. As we consider
here specifically slowly rotating pulsars, in contrast to rapidly 
rotating young pulsars such as the Crab pulsar, pair production 
has little effect on the gap electrodynamics and thus, one can treat
$w_g$ as a constant.

Assume that the outer gap is located along the last open field line starting
from the null surface at a radius $\eta_{\NS}$ and consider acceleration 
of an electron toward the star. Possible extrinsic sources of charged particles are discussed 
in Sec. 3.3. When curvature radiation is the dominant energy loss process,
an assumption used in the usual outer gap models \citep{cetal86,r96}, 
the radiation-reaction limited Lorentz factor is
\begin{eqnarray}
\gamma_c&=&2^{1/2}\left (w_g\gamma_m\eta_{\NS}\right)^{1/4}\left({R_{LC}\over r_e}\right)^{1/4}
\nonumber\\
&\approx&4.1\times10^7w^{1/4}_g\eta^{1/4}_{\NS,0.5}\left({P\over1.5\,{\rm s}}\right)^{-1/4}
\left({B_s\over10^9\,{\rm T}}\right)^{1/4},
\label{eq:gamc}
\end{eqnarray}
where $\gamma_m=e\Phi_m/m_ec^2\approx3.2\times10^7(B_s/10^9\,{\rm T})(P/2\,{\rm s})^{-2}$, 
$\eta_{\NS}=0.5\eta_{\NS,0.5}$, the curvature radius at $\eta_{\NS}$ is assumed to be 
$R_c=(4/3)\eta^{1/2}_{\NS}R_{LC}(1-3\eta_{\NS}/4)^{3/2}/(1-\eta_{\NS}/2)$, 
and $r_e\approx2.8\times10^{-15}\,\rm m$ is the classical radius of the electron.
The characteristic energy (in $m_ec^2$) of curvature photons emitted by these electrons is
\begin{equation}
\epsilon_c={3\over2}\left({\lambda_c\over R_c}\right)\gamma^3\approx
349\,\eta^{1/4}_{\NS,0.5}\left({P\over1.5\,{\rm s}}\right)^{-7/4}\left({B_s\over10^9\,{\rm T}}\right)^{3/4},
\label{eq:ec}
\end{equation}
where $\gamma={\rm min}\{\gamma_c,w_g\gamma_m\}$, the approximation 
is obtained with $\gamma=\gamma_c$, and
$\lambda_c=\hbar/m_ec\approx 3.9\times10^{-13}\,\rm m$ is the Compton wavelength divided 
by $2\pi$. One may estimate the pulsar period range where the acceleration is limited
by curvature radiation:
\begin{equation}
P<1.8\,w_g^{4/7}\eta^{-1/7}_{\NS,0.5}\left({B_s\over10^9\,{\rm T}}\right)^{3/7}\,{\rm s}.
\label{eq:Pcondition}
\end{equation}
The probability of a photon being converted to a pair through collision with 
thermal photons is tiny, and hence the bulk of curvature photons escape as a 
MeV gamma-ray flux with a luminosity
$L_\gamma\approx \epsilon_c N_Lc\pi (\eta_{\NS}R_{LC})^2m_ec^2\sim 10^{19}\,
{\rm J}\,{\rm s}^{-1}$. For slowly rotating pulsars this flux may be too low to be detectable.

For pulsars with rotation periods that do not satisfy (\ref{eq:Pcondition}),
primary particles may lose energy principally through inverse Compton scattering (ICS)
on thermal radiation from the star's surface. For a surface temperature 
$T_s$, the number density of thermal photons 
with a characteristic energy $k_BT_s$ is
\begin{eqnarray}
n_{ph}&\approx& {2\over\pi^2}\left({\Theta_T\over\lambda_c}\right)^3
\left({R_0\over \eta R_{LC}}\right)^2
\nonumber\\
&\approx& 1.3\times10^{17}\,\eta^{-2}\left({T_s\over10^6\,{\rm K}}\right)^3
\left({P\over2\,{\rm s}}\right)^{-2}\,{\rm m}^{-3},
\label{eq:nph}
\end{eqnarray}
where $\Theta_T=k_BT_s/m_ec^2\approx 1.7\times10^{-4}(T_s/10^6\,{\rm K})$.
Since the thermal photons propagate radially, backward moving electrons 
undergo head-on scattering on these photons. In the electron rest frame,
the thermal photon energy is $\gamma \Theta_T\gg 1$, implying that
the scattering is in the Klein-Nishina (KN) regime. The energy loss rate is 
\begin{equation}
-{d\gamma\over dt}\sim {c\sigma_T\over16\lambda^3_c}\,
{\theta^4_d\Theta^2_T\over\eta^2}\, \ln\left(2\gamma\Theta_T \right),
\end{equation}
with $\sigma_T$ the Thomson cross section.

\begin{table*}
\centering
\begin{minipage}{13cm}
\caption{The magnetic field $B_L$ (at the LC), $N_L=N_{GJ}\theta^6_d$,
synchrotron decay time $\tau_d$ times $\gamma$ and charge loading time $\tau_L$. The magnetic
field on the polar cap surface is obtained from $B_s=6.4\times10^{15}(P\dot{P})^{1/2}$
\citep{um95}, twice the value quoted by \citet{metal06}. The time
$\tau_0$ is calculated at $0.5R_{LC}$.}
  \begin{tabular}{@{}lcccccr@{}}
  \hline
 RRATS/Pulsars & $P$ (sec) & $B_s$ ($10^9\,{\rm T}$) & $B_L$ ($10^{-4}\,{\rm T}$) & $N_L$ (${\rm
m}^{-3}$) & $\tau_0$ (sec) & $\tau_L$ (sec)  \\
 \hline
   ${\rm J}1317-5759$  & 2.64 & 1.2 & 5.8 & $7.8\times10^4$ & $2.7\times10^5$ & $1.5\times10^5$\\
   ${\rm J}1819-1458$  & 4.26 & 10  & 11.8 & $9.8\times10^4$ & $6.5\times10^4$ & $93.7$ \\
   ${\rm J}1913+1333$  & 0.92 & 0.54 & 64 & $2.4\times10^6$ & $2.3\times10^3$ & 10.8\\
   Typical magnetars   & 7 & $10^2$ & 18 & $2.5\times10^5$ & $1.3\times10^4$ & 34\\
   The Crab pulsar     & 0.033 & $0.6$ & $1.3\times10^6$ & $9.6\times10^{11}$ &  $4\times10^{-6}$
& - \\
   PSR1957+21         & 0.0015 & $8\times10^{-5}$ & $2.2\times10^6$ & $4.5\times10^{13}$ &
$2\times10^{-6}$ & - \\
\hline
\end{tabular}
\end{minipage}
\end{table*}

\subsection{Pair creation}

The only viable channel for pair creation far from the star's surface
is through photon-photon collision. The primary photons can be
produced from curvature radiation or ICS.
First, consider pair production due to curvature radiation.
The number of curvature photons with energy $\epsilon>\epsilon_c$ emitted 
by a primary electron over a distance $\Delta\varrho R_{LC}$ is written as
\begin{eqnarray}
n_c &\approx&
{3^{3/2}\alpha_f\over 8\pi}\,{
\Delta\varrho\gamma\over\eta^{1/2}_{\NS}}I(\epsilon/\epsilon_c),\\
I(\epsilon/\epsilon_c)&=&\int^\infty_{\epsilon/\epsilon_c}
dy\int^\infty_ydt\,{\rm K}_{5/3}(t),
\end{eqnarray}
where $\Delta\varrho<1$ is the thickness (in units of $R_{LC}$) of the shell region,
$\alpha_f\approx1/137$ is the fine constant and ${\rm K}_n(t)$ is the modified Bessel function.
The number of curvature photons emitted decreases rapidly with increasing $\epsilon/\epsilon_c$.
One has $n_c\approx 6.4\times10^3$ for $\epsilon\geq\epsilon_c$ and
$n_c\approx 1.7\times10^3$ for $\epsilon\geq2\epsilon_c$ and $n_c\approx 2\times10^2$ for
$\epsilon\geq 4\epsilon_c$, with $\Delta\varrho=0.2$,
$\eta_{\NS}=0.5$, and $\gamma=3.2\times10^7$. The number of pairs per electron produced 
near the LC, i.e., the multiplicity denoted by $M_L$, is estimated 
as $M_L\approx n_cn_{ph}\sigma_{\gamma\gamma}R_{LC}\Delta\varrho$ where 
$\sigma_{\gamma\gamma} \approx 2\times10^{-29}\,{\rm m}^2$
is the cross section of pair production via photon-photon collision.
Since pairs are produced on the Wien tail of the thermal spectrum,
in stead of (\ref{eq:nph}), the thermal photon number density is written as
$n_{ph}\approx(2/\pi^2\eta^2)(\Theta_T/\lambda_c)^3\xi_W$,
where $\xi_W=(\epsilon_{th}/\Theta_T)^2\exp(-\epsilon_{th}/\Theta_T)$
is the parameter characterizing the Wien tail ($\epsilon_{ph}\gg\Theta_T$)
where thermal photons satisfy the pair production threshold,
\begin{equation}
\epsilon_{ph}\geq\epsilon_{th}=
{2\over(1-\cos\theta)\epsilon_c},
\label{eq:threshold}
\end{equation}
where $\theta$ is the angle between the pair-producing gamma-ray and thermal 
photon. One has $\xi_W\approx 0.06$ for $\epsilon=4\epsilon_c$, $\theta=\pi/3$ and 
the numbers given in (\ref{eq:ec}).
Assuming that (\ref{eq:Pcondition}) is satisfied, one obtains
\begin{eqnarray}
M_L&\approx& 4.3\times10^{-3}w^{1/4}_g
\left({T_s\over10^6\,{\rm K}}\right)^3\left({
P\over1.5\,{\rm s}}\right)^{-5/4}
\left({B_s\over10^9\,{\rm T}}\right)^{1/4}\nonumber\\
&&\times
\left({\xi_W\over0.06}\right)
{\Delta\varrho_{0.2}\over\eta_{\NS,0.5}},
\label{eq:M}
\end{eqnarray} 
with $\Delta \varrho_{0.2}=0.2\Delta\varrho$. 
The multiplicity increases rapidly with decreasing pulsar period. 
As examples, from Eq (\ref{eq:M}) one obtains
$M_L\approx0.027$ for J1913+1333 with $T_s=10^6\,\rm K$ and
$M_L\approx0.015$ for J1819$-$1458 with $T_s=1.4\times10^6\,\rm K$, where $\eta_{\NS}=0.5$
and $\Delta\varrho=0.2$ is assumed.

Pair production due to ICS becomes significant if the curvature photon energy 
is too low to satisfy the threshold condition (\ref{eq:threshold}).
Since the scattered photon energy is $\epsilon_\gamma\sim \gamma$
in the KN regime, the production rate of the scattered photons with
energy $\epsilon_\gamma$ is $dn_\ICS/dt\approx
-\gamma^{-1}d\gamma/dt$. The number of photons emitted through ICS over a distance 
$\eta_{\NS}R_{LC}$ is
\begin{equation}
n_\ICS\approx 
{\sigma_TR_{LC}\over16\lambda^3_c}\,
{\theta^4_d\Theta^2_T\over\eta_{\NS}}\,
{\ln(2\gamma_m\Theta_T)\over\gamma_m}.
\end{equation}
Similarly, one has $M_L\approx n_\ICS n_{ph}\sigma_{\gamma\gamma}R_{LC}\Delta\varrho$,
which takes the following numerical form
\begin{equation}
M_L\approx 2.1\times10^{-4}\left({T_s\over10^6\,{\rm K}}\right)^5\left({B_s\over10^9\,{\rm T}}
\right)^{-1}{\Delta\varrho_{0.2}\over\eta^3_{\NS,0.5}}.
\end{equation}
Pair production is strongly dependent on the surface temperature but
insensitive to the pulsar period. 

In the outer gap models, the primary particles are assumed to be created 
in a pair cascade in the gap \citep{cetal86,r96} and the primary flux is limited to
a value near the GJ flux. To allow a substantial energy transfer from primaries to pairs,
as this is the case for high-energy pulsars, one must have the pair multiplicity in the vicinity of the
gap to be at least one, i.e., $M_g\geq1$, with $M_g=1$ defining a deathline where the gap becomes inactive. 
For parameters appropriate for RRATs and magnetars, one has $M_g\ll1$ except for $\alpha\to\pi/2$. 
Thus, one can ignore the feedback effect by pair production on the accelerating 
electric field in the gap.

\subsection{Source of primary particles}

Low-level accretion of neutral grains from a dust disk or the ISM can provide 
extrinsic charged particles. The accreted neutral material can be ionized inside the 
pulsar magnetosphere by the thermal radiation from the star's surface and
particles produced from ionization are then channeled to the outer gap
or trapped in the CFLR. The maximum flux that can be tolerated 
in the gap is about the GJ flux $cN_L/\eta^3_{\NS}$ with the gap location assumed to be
along the last open field lines starting from the null surface at the radius
$\eta_{\NS}$. It is unlikely that the supply of charged particles 
matches exactly the GJ flux and the acceleration may become oscillatory so that the
net flux is maintained at about the GJ level. 
For the case of accretion of the ISM dust grains,
the accretion rate is $\dot{M}\sim (GM)^2\rho_{ISM}/v^3\sim 
10^4\,{\rm kg}\,{\rm s}^{-1}$, where $G$ is the gravitation constant,
$M$ is the neutron star's mass,
and $v$ is the pulsar velocity, $\rho_{ISM}$ is
the ISM density. Dust grains of small size ($<0.1\mu\,\rm m$) may not survive 
sputtering by protons when crossing the bow shock and those with size
$\sim0.1\mu\,\rm m$ have a better chance to reach the LC \citep{c85}. 
Assuming that flux rate $F_e$ of electrons into the gap is $10\%$ of this flux rate,
one estimates $F_e=c\dot{N}_a\sim 0.1(c/2\pi R^2_{LC})(\dot{M}/m_p) 
\approx5\times10^{13}\,{\rm m}^{-2}\,{\rm s}^{-1}$. Here we assume
$v=100\,{\rm km}\,{\rm s}^{-1}$ and $\rho_{ISM}=10^{-21}\,{\rm kg}\,{\rm m}^{-3}$.
One finds this flux is considerably larger than the GJ flux $cN_L\sim 2.4\times10^{13}
\,{\rm m}^{-2}\,{\rm s}^{-1}$. 
The accelerating electric field changes sign if the 
flux exceeds the GJ flux \citep{hs02}. 
Although a divergent solution for $E_\parallel$ is possible \citep{hs02}, here we
assume that the maximum accelerating potential is limited by (\ref{eq:Ep})
determined by the vacuum potential drop.

For pulsars with $T_s>10^5\,\rm K$, the ISM grains may not survive
disintegration \citep{c85} and in this case the disk is the main source of 
the accreted neutral matter. The disk model was discussed in detail
by \citet{cs06}. The recent discovery of a fall-back disk
around a magnetar lends support for the existence of such disks
around long-period pulsars. 

\subsection{Particle loading in the CFLR}

Despite the slow rate of pair production, a significant number of
particles can be accumulated in the CFLR over a time $\gg P$.
A schematic of pair injection by a latent outer gap is shown in Figure \ref{fig:rrat}.

One may estimate the time required for the plasma density to increase to the
GJ density in the trapping region.
Assume a particle flux in the gap is $cN_L/\eta^3_{\NS}$.
Acceleration of these particles can inject pairs into the CFLR at a rate 
$\pi MN_Lc\delta_s R^2_{LC}/\eta_{\NS}$ with the gap's cross section area approximated by
$\pi \delta_s\eta^2_{\NS}R^2_{LC}$ (where $\delta_s<1$). A continuous injection of pairs can lead
to build-up of the plasma density in the trapping region. The time
for the density to reach the GJ density $N_L$ is 
\begin{equation}
\tau_L={\Delta\varrho P\over2\pi M\delta_s}.
\label{eq:tauL}
\end{equation}
We refer to (\ref{eq:tauL}) as the pair loading time.
Numerical examples of $\tau_L$ are shown in table 1, where we assume
$\Delta\varrho=0.2$, $\eta_{\NS}=0.5$, and $\theta=\pi/3$. The surface
temperature is assumed to be $T_s=10^6\,{\rm K}$ for J1913+1333, 
$T_s=1.4\times10^6\,{\rm K}$  for J1819-1458 \citep{retal06} and 
$T_s=1.5\times10^6\,{\rm K}$ for a magnetar. One has
$\tau_c\gg \tau_L$ for these three pulsars. For J1317-5759, one finds
$\tau_L\sim 1.5\times10^5\,\rm s$ for $\theta=\pi/2$, $T_s=1.5\times10^6\,\rm K$ and
$\eta_{\NS}=0.2$.

Alternatively, the plasma can be injected near the star through
photon decay in the magnetic field \citep{wetal98}. Such case was also considered 
by \cite{rg05} for
the double pulsar system. In their model, they considered pitch angle increase
due to absorption of the coherent radio waves from the other pulsar in
the system. 

\section{Injection through ionization}

Radiation belts can form directly through evaporation and 
ionization of neutral grains accreted from
a disk or the ISM. One can show that the density of trapped plasmas
can easily exceed the local GJ density with a slow accretion rate.
For example, to accumulate a plasma with the GJ density over a time $\tau_L$, 
a rate as low as $cN_L/\tau_L$ would be sufficient. 
The trapped particles can be accelerated to relativistic energy through 
either cross field diffusion toward the star or interactions with 
waves near the cyclotron resonance. Both processes are believed to be responsible
for particle acceleration in the Earth's van Allen belts where electrons are
accelerated up to 10 MeV energy \citep{hetal05}. 
In the cross field diffusion, extremely low-frequency disturbances (ELFWs)
in resonance with the particle drift frequency (around the star)
can force particles to diffusion across the field lines inward
(toward the star) where the magnetic field increases. Conservation of both
the first adiabatic invariant $p^2_\perp/B$ and the second adiabatic invariant
$\oint ds p_\parallel$ (bouncing back and forth between two mirror points)
implies an increase in both $p_\perp\propto 1/r^3$ and  $p_\parallel\propto 1/r$;
thus, $p$ increases as $\tan\alpha=p_\perp/p_\parallel\propto 1/r^{1/2}$ increases with decreasing
$r$ (inward), that is, the particles are accelerated. For the Earth's radiation belts, 
ELFWs are generated due to the large scale spatial disturbances
by the solar wind. In the pulsar case, the most plausible candidate for such low
frequency waves is Alfv\'en waves (cf. Sec. 6).

\section{Stability of the trapped plasma}

\begin{figure}
\includegraphics[width=8cm]{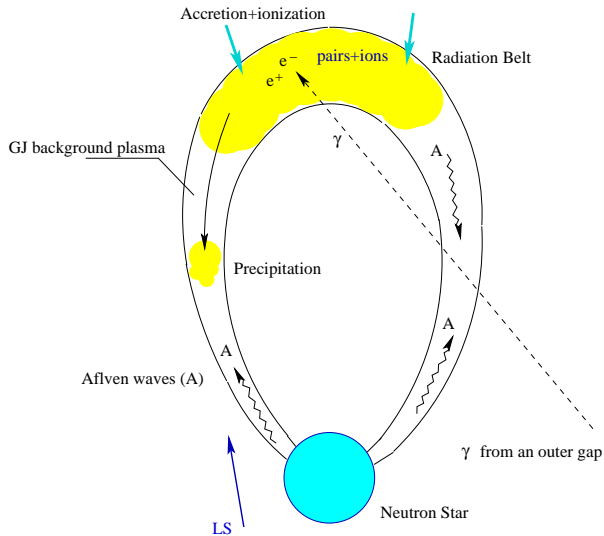}
\caption{Disruption of the trapping region and particle
precipitation in the CFLR. Alfv\'en waves ($A$) emitted from the surface can bounce
back and forth in the closed magnetic flux tube interacting with the trapped particles.
LS is the direction in which an observer would see the bursts arising from
streaming instability as a result of downward moving particles. In contrast,
bursts arising from cyclotron maser at reflection points can only be seen in the 
direction approximately perpendicular to the field lines.}
\label{fig:prb}
\end{figure}

The pulsar radiation belts have many similarities to planetary radiation belts.
For example, in both cases the stability of the trapped plasma can be strongly 
affected by change of the global magnetic field structure, i.e. distortion of the 
`magnetic bottle', or strong pitch-angle scattering of particles on waves. For the Earth's van
Allen belts, such activities are due to the disturbances from the solar wind,
while activities in the pulsar radiation belts are powered by the rotational energy 
or magnetic energy. In magnetars, strong starquakes can lead to significant distortion of magnetic field lines,
but such major events may not be common for normal pulsars.
Here we only discuss low-level starquakes or neutron star oscillations (cf. Fig \ref{fig:prb}), which
can lead to pitch-angle scattering by Alfv\'en waves generated at the surface.

\subsection{Pitch-angle scattering}

A formal theory for diffusion of particles in momentum space is the quasilinear 
diffusion formalism in which the processes arising from a feedback from 
growth of linear waves are determined by diffusion coefficients. In momentum space, such diffusion can
be separated into diffusion in pitch angle and in $p$. The general formalism is outlined in Appendix
B.  Here we consider specifically the pitch angle diffusion coefficient
$D_{\alpha\alpha}$ (cf. Appendix). Quasilinear diffusion occurs effectively through resonant
wave-particle interactions. The most widely-discussed process in the context 
of cosmic ray confinement is the scattering due to cyclotron resonance 
$\omega-k_\parallel v_\parallel -s\Omega_e/\gamma=0$ with $s=\pm1$ \citep{m80}. In this case,
the wave fluctuations can be regarded as predominantly magnetic and as the zeroth order
approximation in the expansion on $\omega/kc\ll1$, the process can 
be viewed as elastic scattering of particles by magnetic
fluctuations, causing change of their motion direction.
Since the frequency of waves generated from starquakes \citep{betal89} or
stellar oscillations \citep{metal88} is low, about kHz as compared to 
the cyclotron frequency $>10\,\rm MHz$, the cyclotron resonance condition is generally not satisfied
except for the cases where these waves can cascade into high-frequency waves, 
e.g., through three-wave interactions. Here we only consider
pitch-angle scattering due to the Cerenkov resonance. 

Unlike wave-particle interactions in the cyclotron resonance, the Cerenkov resonance 
only changes $p_\parallel$ not $p_\perp$. In general, it leads to an increase 
in $p_\parallel$, corresponding to a decreasing pitch angle $\alpha$.
There are two relevant low-frequency wave modes: the $X$ mode and 
the $LO$, corresponding respectively to the fast mode 
and the Alfv\'en mode in the MHD.
Pitch angle diffusion due to the fast mode waves in Cerenkov resonance was
discussed by~\citet{sm98}. Since there is a nonzero $\delta \bB$ along the mean magnetic field,
fast particles in the Cerenkov resonance bounce back and forth between
two successive magnetic compressions (acting as two magnetic mirrors) in the wave frame. On average, 
particles gain energy causing an increase in $p_\parallel$. This process, also called 
transit-time damping, works for a low  Alfv\'en speed $v_A/v\ll1$, a condition not satisfied for pulsars.
So, instead we consider a low frequency $LO$ mode, which we refer to as the Alfv\'en wave.
In kinetic theory in the limit $v_A\to c$, which is appropriate for pulsar
magnetopheric plasmas, Alfv\'en waves have a nonzero $\delta \bE$ along the magnetic field
\citep{ab86,mg99} and particles can be accelerated through the 
Cerenkov resonance \citep{vetal85}.
The scattering time $t_\alpha=1/D_{\alpha\alpha}$ may be
written in terms of the ratio of the magnetic energy density 
$U_B=\delta B^2_A/2\mu_0$ of the wave to the plasma kinetic
energy density $U_p=m_ec^2\langle\gamma\rangle$ (Appendix)
\begin{equation}
t_\alpha\approx
{1\over2\pi\omega}\left({U_p\over U_B}\right)
\left({\omega_p\over \omega}\right)^2{1\over
\sin^2\theta\,\sin^2\alpha\,\cos\alpha}.
\label{eq:ta1}
\end{equation}
Considering a plasma consisting of electrons and positrons,
for $U_B\sim U_p$, $\omega=0.1\omega_p$, $\theta=0.3$ and $\alpha=\pi/4$, one 
estimates $t_\alpha\approx 1\,\rm s$. One concludes that
diffusion due to particles in Cerenkov resonance 
with a low frequency Alfv\'en wave can effectively 
transfer particles to small pitch angles.

It is appropriate to comment here that growth or damping of Alfv\'en waves through 
the Cerenkov resonance is generally not effective because the parallel component of
the wave polarization is rather small, $e_\parallel\sim(\omega/\omega_p)^2\ll1$.  
Thus, quasilinear diffusion as a result of such linear wave growth is not significant, limited
by $U_B/U_p\ll1$ (cf. Eq. \ref{eq:ta1}). This has been the main reason
that such two-stage diffusion mechanism is not favored for the interpretation of cosmic-ray propagation, which,
as inferred from observations, is subject to strong-pitch angle scattering in the 
ISM \citep{m80}. However, such limit is not applicable in our case as Alfv\'en waves are 
assumed to be generated externally, not through the Cerenkov resonance, with magnetic 
energy density comparable with $U_p$; thus, efficient diffusion occurs 
even for $(\omega/\omega_p)^2\ll1$.

\subsection{Precipitation}

Disturbances to the radiation belt can be catastrophic if the wave that propagates
to the region is in the form of short bursts producing 
$U_B\sim U_p$ in the trapping region. As shown from 
(\ref{eq:ta1}), the scattering time can be as short as seconds.
The requirement of $U_B\sim U_p$ is rather modest.
For example, for $\langle\gamma\rangle=5$, one has 
$U_p\approx 5\times10^{-7}\,{\rm J}\,{\rm m}^{-3}$, giving
a luminosity $\Delta\varrho R^3_{LC}U_p/\Delta t\approx 10^{16}\Delta\varrho_{0.1}
(1\,{\rm s}/\Delta t)\,\rm W$ where $\Delta t$ is the pulse duration.
As this luminosity is minuscule compared with transient emission in magnetars (typically 
$10^{25}\,{\rm W}$), these waves are not significant in producing high-energy emission.
Strong pitch angle diffusion channels most particles into the much smaller
loss cone through which particles can reach a region close to the star
where their synchrotron decay time is short. This can cause a sudden intense 
precipitation of particles on a time scale shorter than the pulsar period.
One may estimate the required wave amplitude in terms of magnetic fluctuations
as $\delta B_A/B_L\sim (U_p/U_{BL})^{1/2}\approx 2\times10^{-4}$, 
implying low intensity quakes or oscillations would be sufficient for triggering
disruption of the trapped plasma. 

If the triggering mechanism is starquakes, the reoccurrence time of such particle storms
is essentially the reoccurrence time of the quakes. The physics of neutron star
quakes or oscillations is not well understood. A possibility is
that a strong toroidal field exists in the crust underneath the surface
and its stress can be released as quakes transfering the magnetic energy to
the magnetosphere in Alfv\'en waves \citep{betal89,td95}. One expects that such starquakes are
relatively more frequent for a high-field, young pulsar. If 
generation of low-intensity Alfv\'en waves is so frequent that 
the reoccurrence time is much shorter than the pulsar period, trapped plasmas do not 
have enough time to accumulate and radiation belts may not form. This may be the case
for magnetars with active transient emission. 

\section{Application to RRATs}

An important consequence of the formation of pulsar radiation belts is
transient emission---radio bursts that can be produced as particles precipitate 
and such transient emission may be relevant to RRATs. 

\subsection{Coherent emission}

RRATs are bright radio bursts with a flux density up to a few Jy at 1.4 GHz. 
As for normal radio pulses, the emission mechanism must be coherent. 
Since there is no widely accepted mechanism for the pulsar radio emission, 
it is expected that many uncertainties would remain in 
identifying the specific emission mechanism for the transient case. Nonetheless,
in the following we consider two possibilities 
in particular: two-stream instability and cyclotron/synchrotron maser. 

\subsubsection*{Two-stream instability}

The two-stream instability has been discussed extensively 
as a possible mechanism for pulsar radio emission, but
the growth rate is not sufficient for the instability to develop in the polar
region, mainly because of the very high Lorentz factor of the primary particle
beam \citep{u87}. It is shown here that the plasma conditions
created during the particle precipitation is favorable for such an instability to
occur.

Two-stream instability can develop when particles traverse a corotating 
background plasma which is assumed to be stationary. Assuming the precipitating 
plasma density is $2M_L$ times the local GJ density $N_L\eta^{-3}$ and the spread
in the particle momentum distribution is ignored (this spread can 
reduce the growth rate, cf discussion below). Consider 
wave growth due to the Cerenkov resonance $\omega\approx k_\parallel v_\parallel$.
The characteristic frequency is determined by the denser plasma.
For $2M_L\ll1$, the characteristic frequency is the frequency
$\omega_p=(e^2N_L/\varepsilon_0m_e)^{1/2}\eta^{-3/2}$ of the stationary background, corotating plasma, with 
the backflowing plasma regarded as a weak beam satisfying 
the resonance condition. The growth rate is then estimated to be
\begin{equation}
\Gamma\approx {\sqrt{3}\over2}\left({M^{1/3}_L\over2\gamma}\right)\,\omega.
\label{eq:Gamma2}
\end{equation}
The growth rate reaches maximum at $\gamma\to\gamma_c$ in the limit $p_\perp\to 1$.
For $2M_L\gg1$, the growth rate can be estimated in the rest frame of the 
backflowing plasma assuming that the corotating background plasma is a beam in Cerenkov resonance.
One finds
\begin{equation}
\Gamma\sim {\sqrt{3}\over2}{\omega\over(4M_L)^{1/3}\gamma},
\label{eq:Gamma3}
\end{equation}
where $\omega=\omega_p(2M_L\gamma)^{1/2}$ is the characteristic frequency seen in the
pulsar frame. 

Note that for two-stream instability in the polar cap model, the energetic particle
beam is usually identified as those primary particles from the PC, with 
a Lorentz factor as high as $\sim 10^6-10^7$ and instability is strongly suppressed 
\citep{cr77,u87}. In the model discussed here $\gamma_c$ is about a few and 
the growth rate (\ref{eq:Gamma2}) or (\ref{eq:Gamma3}) can be quite fast. 
For $\gamma\to\gamma_c=2$, $M_L=0.1$, (\ref{eq:Gamma2}) gives $\Gamma/\omega\approx 0.1$.
Although the growth rate can be reduced due to a spread in the 
particle momentum distribution, it seems that the instability can develop rapidly.
It should be pointed out here that radio emission from two-stream instability is
a two-stage process. The instability can generate Langmuir 
waves which can in turn be converted to electromagnetic waves and at the same time facilitate
the particle precipitation. 

\subsubsection*{Cyclotron/synchrotron maser}

Since particles with small pitch angles escape through the loss cone,
particle precipitation creates an inversion in the pitch angle distribution
of the upward flowing particles, driving cyclotron maser \citep{md82,hetal82} or synchrotron maser
\citep{zs72,y90,ha91}. This mechanism was discussed widely in interpretation of 
planetary radio bursts \citep{m93,eetal00} and solar bursts \citep{md82}. 
Synchrotron maser is applicable in the relativistic regime
$p_\perp>1$, while cyclotron maser may be the more relevant when such inversion occurs close to the star
during the precipitation when their gyration is in the nonrelativistic regime.
The relevant radio wave can grow when the frequency satisfies the cyclotron resonance condition.
For cyclotron resonance at the fundmental harmonic $\omega-\Omega_e/\gamma-k_\parallel v_\parallel=0$,
the growth rate is given by \citep{md82}
\begin{equation}
\Gamma\approx {N_b\over N_L}\left({\omega_p\over\Omega_e}\right)^2{c^2\over v^2_0},
\end{equation}
where $N_b$ is the number density of particles in cyclotron resonance,
and $v_0$ is the velocity of particles in cyclotron resonance. For a typical 
pulsar with $P=1.5\,\rm s$ and $B_s=10^9\,\rm T$, this condition can be
satisfied at a radius $0.1R_{LC}$. It is worth noting that the related but different 
process---cyclotron instability due to the anomalous doppler effect, which has been considered 
in the literature for pulsar radio emission \citep{ketal91}, is not applicable here; it requires 
an energetic beam ($\gamma\sim 10^6-10^7$) of particles that satisfy the anomalous doppler condition.

\subsection{Beaming properties}

Bursts produced from particle precipitation are characterized by sharp spikes.
In the case of two-stream instability, since emission arising from the instability
is confined to a thin shell region in the CFLR near the last open field lines, one expects the beaming direction
of the radio emission to be similar to that of reverse emission in the conventional
polar cap models \citep{detal05,zetal07}. Consider a trapping region has 
a radial range from $R_{LC}(1-\Delta\varrho)$ to $R_{LC}$. The inward tangent angles at $\eta$ are
respectively given by $\psi_2$ and $\psi_1$. One has a conal structure with an
angular thickness as $\Delta\psi=\psi_2-\psi_1
\approx(3/4)\Delta\varrho\eta^{1/2}$ for $\Delta\varrho\ll1$.
For the emission to be in radio, the emission radius has to be $\eta\sim 10^{-2}$, giving
$\Delta\psi\approx 0.05$ for $\Delta\varrho=0.1$. Thus, the model predicts a rather
narrow burst profile. Since particle precipitation and subsequent burst emission 
can be highly localized and occurs over a large range of altitudes, such
bursts can appear on a wide range of pulsar phase. 

For cyclotron maser, the beam width is  $\Delta\theta\sim v_0/c\ll 1$ \citep{hetal82};
thus, the maser emission can produce narrow spiky profiles. Since cyclotron maser 
emission occurs near particle reflection, the emission is beamed at a large
angle to the field lines. This beaming feature is very different from
the usual relativistic beaming ($1/\gamma\ll1$) along the field lines.
Such unsual beaming feature can be tested observationally 
if both bursts and weak emission (i.e., the usual pulsed emission) 
can be detected and relative phase between the two components can be determined.

\subsection{Propagation}

Propagation of radio emission in the CFLR is constrained by 
induced scattering by electrostatic waves excited by the radio emission 
itself \citep{lm06}. Assuming a brightness temperature $T_b$ and a beaming solid angle
$\Delta\Omega$, one may define a radius of an opaque sphere \citep{lm06},
\begin{eqnarray}
\eta_{cr}&\approx& 3\times10^{-3}
\left({\Delta\Omega T_b\over10^{24}\,{\rm K}}\right)^{2/7}
\left({\nu\over1.4\,{\rm GHz}}\right)^{-2/7}
\left({P\over2\,{\rm s}}\right)^{-10/7}\nonumber\\
&&\times \left({B\over10^9\,{\rm T}}\right)^{3/7},
\label{eq:rcr}
\end{eqnarray}
such that the wave with a frequency $\nu$ can only propagate outside the
sphere $\eta>\eta_{cr}$. Here we assume that the plasma density in the propagation 
path is the local GJ density. 

Although $\Delta\Omega T_b$ is not well constrained, the condition (\ref{eq:rcr})
is quite robust. For an observed flux density $S^{\rm obs}_\nu$ and a pulsar distance $D_L$,
the brightness temperature times the beaming solid angle can be estimated from
\begin{eqnarray}
\Delta\Omega T_b&\approx&
2.3\times10^{23}\,{\rm K}
\left({\Delta s\over 10^6\,{\rm m}}\right)^{-2}
\left({1.4\,{\rm GHz}\over\nu}\right)^2\nonumber\\
&&\times\left({D_L\over3\,{\rm kpc}}\right)^2
\left({S^{\rm obs}_\nu\over1\,{\rm Jy}}\right),
\end{eqnarray}
where $\Delta s$ is the distance between the scattering region and the emission region.
In the case of inward emission due to two-stream instability,
(\ref{eq:rcr}) gives a lower limit to the emission radius; for the 
emission to be potentially visible, the emission radius must be $\eta>4\eta_{cr}$
\citep{lm06}. 

\section{Conclusions and discussion}

We consider possible existence of transient radiation belts in
the magnetospheres of pulsars with low LC magnetic fields. It is suggested
that particle acceleration in latent outer gaps can lead to creation of pairs with
large pitch angles in the CFLRs near the LC where they are trapped due to the 
magnetic mirror effect. In the trapping regions where the magnetic fields are weak 
($\sim 10^{-4}\,\rm T$), particles radiate away their perpendicular energy over a time
much longer than a typical pulsar period. 
Thus, the plasma density can build up over a synchrotron cooling time, which is much longer than
the pulsar period. Similar to the planetary radiation belts, pulsar radiation belts can be 
disrupted due to disturbances of waves propagating into the regions. Catastrophic 
disruption of these regions may lead to intense particle precipitation, which in turn 
generates radio bursts that can be seen as radio transients. The main triggering mechanism is 
scattering of trapped particles into a loss cone as a result of interactions with 
waves generated in plasma instabilities in the trapping regions or 
Alfv\'en waves emitted at the surface due to low-intensity starquakes or oscillations. 
The plasma that rushes through the background corotating plasma can produce bursts of coherent radio 
emission through streaming instability or cylcotron/synchrotron maser. Existence of such 
radiation belts, if observationally confirmed, would imply that some of the bursty phenomena
of pulsar radio emission may be due to particle precipitation. In particular, we suggest 
that such transient radio emission may be seen as RRATs.

A notable feature of our model, as applied to RRATs, is that in comparison with conventional polar cap models,
the transient emission mechanism considered here is less constrained by the pair production 
efficiency; it predicts that pulsars below the `deathline' can 
still produce radio bursts. One of the major problems with polar cap models is the low
efficiency of pair production. Most RRATs have long periods and for pulsars with such long periods,
there is insufficient supply of pairs needed for coherent emission. Although an 
outer gap can supply additional pairs through acceleration of charged
particles supplied externally \citep{c85,rc88,cs06}, the pair production
efficiency is too low to produce a substantial downflow of pairs; the only possible
location for reverse emission in the polar regions is near or in the polar gap
where a downward pair cascade can occur; for example, externally-supplied
particles flush the polar gap resulting in a downward transient pair cascade
followed by quenching of the gap \citep{cs06}. However, such a scenario would predict
an emission region very close to the star and any reverse emission
may be eclipsed by either the star or a dense plasma near the surface.
In our model, since particle precipitation occurs in the CFLR,
it does not quench the polar gap, allowing
the possibility of observing both burst and normal radio emission. 

In application to magnetars with active high-energy transient emission, one expects 
low-intensity quakes to occur much more frequently than in normal pulsars \citep{td95}. Such frequent 
disruption may prevent a radiation belt forming in a magnetar magnetosphere, even though
the synchrotron decay time is relatively long near the LC. This scenario seems to be supported
by lack of any detection of RRAT-like radio bursts from the known magnetars.  
Although transient radio emission was recently detected from the magnetar 
XTE J1810$-$197, polarization study suggests that the emission geometry
resembles that of young pulsars \citep{cetal07}. 

In principle, our model may also apply to old, long-period pulsars. The 
loading time $\tau_L$ can be significantly longer than the synchrotron decay time.
The plasma trapping can be due to {\it in situ} ionization of the accreted neutral grains
in the CFLR. Since the synchrotron decay time is long, for example, $\tau_s\approx 16\,\rm yr$
for PSR J2144$-$3933 with $P=8.5\,\rm s$ and $B_s=4\times10^8\,\rm T$, a nearby long-period pulsar,
accumulation of plasma can occur over a long time (provided that the reoccurrence time 
of the triggering mechanism is also long). Since old pulsars are not energetic,
the predicted radio luminosity is low. As it takes a long time to replenish the trapped plasma, such
transient events can be rather infrequent but may still be detectable provided 
these pulsars are located relatively nearby. 

\section*{Acknowledgements}
We thank Kouichi Hirotani for helpful discussion.

\appendix

\section{Bouncing time}
From the first adiabatic invariant $p^2_\perp/B={\rm const}$ and the energy conservation
$p^2={\rm const}$, one has
\begin{equation}
p^2={B(\eta)\over B(\eta_{i})}p^2_\perp(\eta_{i})+p^2_\parallel(\eta),
\end{equation}
where the relevant quantities are functions of $\eta$.
Assuming $\sin\alpha_{i}=p_{\perp}(\eta_{i})/p$, the time to travel
from the injection radius $\eta_i$ to the reflection radius $\eta_r$ and back to $\eta_i$
can be written in terms of integration along the particle's path $ds$:
\begin{eqnarray}
\Delta t&=&{2\over v}\int{ds\over\left[1-B(\eta)/B(\eta_r)\right]^{1/2}}
\nonumber\\
&\approx& {R_{LC}\over v}\int^{\eta_r}_{\eta_i}\left({4-3\eta\over1-\eta}\right)^{1/2}
{d\eta\over\left[1-(\eta_r/\eta)^3\right]^{1/2}},
\end{eqnarray}
where we assume a particle follows the last closed field line of a dipole field and
$\eta_r<\eta_i$ satisfies the condition $1-[B(\eta_r)/B(\eta_i)]\sin^2\alpha_i=0$.
The integration yields approximately $\Delta t\approx \eta_i P/\pi$.
One may assume without loss
of generality $\eta_i=1$ to show that the bounce time (back and forth between two mirror points)
is $\tau_b=2\Delta t\approx 2P/\pi$. The bounce time is not sensitive to
the pitch angle at the injection provided that the condition $\alpha_i\geq\theta^6_d$ is
satisfied; this condition requires the reflection radius to be larger
than the stellar radius, $\eta_r\geq R_0/R_{LC}$.

\section{Pitch-angle scattering}

We outline quasilinear theory for pitch-angle scattering. Let
$f(p,\alpha)$ be a gyrophase-averaged distribution in the particle 
momentum space. In the quasilinear diffusion formalism one has
\begin{eqnarray}
{df\over dt}
&=& {1\over\sin\alpha}{\partial\over\partial\alpha}\Biggl[\sin\alpha
\left(D_{\alpha\alpha}{\partial \over\partial
\alpha}+{\partial\over\partial\alpha}D_{\alpha p}{\partial \over
\partial p}\right)f\Biggr]\nonumber\\
&&+{1\over p^2}{\partial\over\partial p}\Biggl[
p^2\left(D_{p\alpha}{\partial \over\partial \alpha}+D_{pp}
{\partial \over\partial p}\right)f\Biggr],
\end{eqnarray}
where the diffusion coefficients are written in terms of the scattering
probability $w(s,\bp,\bk)$ and the wave occupation number $N(\bk)$ \citep{m80}:
\begin{equation}
\left[
\begin{array}{c}
D_{\alpha\alpha}\\
D_{\alpha p}\\
D_{pp}
\end{array}
\right]=
\sum_s\int{d\bk\over(2\pi)^3}w(s,\bk,\bp)\, N(\bk)
\left[
\begin{array}{c}
(\Delta\alpha)^2\\
\Delta\alpha \Delta p\\
(\Delta p)^2
\end{array}
\right]
\end{equation}
with $D_{\alpha p}=D_{p\alpha}$, $\Delta\alpha=\hbar(\omega\cos\alpha-k_\parallel
v)/(pv\sin\alpha)$, and $\Delta p=\hbar\omega/v$. The scattering probability is 
\begin{eqnarray}
w(s,\bp,\bk)&=&{2\pi e^2R\over\varepsilon_0\hbar\omega}
\left|\be\cdot\bV(s,\bp,\bk)\right|^2\nonumber\\
&&\times\delta\left(\omega-k_\parallel v_\parallel-
s\Omega_e/\gamma\right),
\end{eqnarray}
\begin{eqnarray}
\bV(s,\bp,\bk)&=&\biggl({1\over2}v_\perp\left[
e^{i\phi}J_{s-1}(z)+e^{-i\delta_e\phi}J_{s+1}(z)\right],
\nonumber\\ 
&& -{1\over2}i\delta_e v_\perp\left[
e^{i\phi}J_{s-1}(z)-e^{-i\delta_e\phi}J_{s+1}(z)\right],\nonumber\\
&&v_\parallel J_s(z)\biggr),
\end{eqnarray}
with $\delta_e=1$ for positrons and $\delta_e=-1$ for electrons,
$R$ is the ratio of the electric to total wave energy density,
$z=k_\perp v_\perp\gamma/\Omega_e$ and $\bk=(k_\perp\cos\phi,k_\perp\sin\phi,k_\parallel)$.
In the following we consider the small $z\ll1$ limit.
The Bessel function has an approximation
$J_s\approx (z/2)^s/\Gamma(s+1)$ for $s\geq0$. This gives 
$J'_0=-J_1$ and $J_{-1}=J_1$.

The polarization of a low-frequency Alfv\'en wave can be written as
\begin{equation}
\be_A\approx(\cos\phi,\sin\phi,e_{A\parallel}),
\end{equation}
where $e_{A\parallel}\approx (\omega/\omega_p)^2(k_\perp/k_\parallel)n$
\citep{ab86,mg99}, and $n$ is the refractive index.
It can easily be verified that $\be_A\cdot\bV=\be_{A\parallel} v_\parallel$ for $s=0$.
Assuming $R=1/2$ and $\sin\theta=k_\perp/k$, one obtains
\begin{equation}
w(0,\bp,\bk)={\pi e^2v^2_\parallel\over\varepsilon_0\hbar\omega}\left(
{\omega\over\omega_p}\right)^4n^2\sin^2\theta\,\delta(\omega-k_\parallel v_\parallel).
\end{equation}
The diffusion coefficient is
\begin{equation}
D_{\alpha\alpha}\approx
2\pi\omega\left({U_B\over U_p}\right)\left({k_\parallel\over\Delta k_\parallel}\right)
\left({n\omega\over\omega_p}\right)^2\sin^2\theta\,\sin^2\alpha\,\cos\alpha,
\end{equation}
where $\omega\approx k_\parallel c$, 
$U_B=\delta B^2_A/2\mu_0$ is the magnetic energy density of 
the wave concerned, and $U_p=m_ec^2\langle\gamma\rangle N_LM_L$ is the 
density of the plasma kinetic energy. It should be emphasized here that the disffusion
in pitch angle is caused by change in $p_\parallel$ not $p_\perp$.

\label{lastpage}

\end{document}